# KNOWM Memristors in a Bridge Synapse delay-based Reservoir Computing system for detection of epileptic seizures

Dawid Przyczyna,[*1,2] Grzegorz Hess[3] and Konrad Szaciłowski[*1]

1.   *AGH University of Science and Technology, Academic Centre for Materials and Nanotechnology,*

     *al. Mickiewicza 30, 30-059 Kraków, Poland*

2.   *AGH University of Science and Technology, Faculty of Physics and Applied Computer Science,*

     *al. Mickiewicza 30, 30-059 Kraków, Poland*

*3 Jagiellonian University, Department of Neurophysiology and Chronobiology, Institute of Zoology and Biomedical Research,*

*ul. Gronostajowa 9, 30-387 Kraków, Poland*

## 1. Abstract

Nanodevices that show the potential for non-linear transformation of electrical signals and various forms of memory can be successfully used in new computational paradigms, such as neuromorphic or reservoir computing (RC). Dedicated hardware implementations based on functional neuromorphic structures significantly reduce energy consumption and/or increase computational capabilities of a given artificial neural network system. Concepts of RC, which as a flexible computational paradigm can be highly inclusive, are often used as a model to describe computations performed *in materia.* With mostly fixed internal structure, solid-state devices, especially memristors, are studied as computational substrates in various RC systems. In this work, we present single-node Echo State Machine (SNESM) RC system based on bridge synapse as a computational substrate (consisting of 4 memristors and a differential amplifier) used for epileptic seizure detection. KNOWM memristors were posed as ideal candidates because of their easy prototyping and reliability of operation. In this account, we present an application of commercially available KNOWM memristors in various neuromorphic applications, from simple analysis of switching and internal dynamics (elucidated form noise spectroscopy and total harmonic distortion analysis) to the classification and recognition of complex time series: epilepsy seizure recognition using a wrist-worn triaxial accelerometer. The results show that the evolution of the signal in a feedback loop helps improve the classification accuracy of the system for that task. The transformation in SNESM changes the correlation and distribution of the complexity parameters of the input signal. In general, there are more differences in the correlation of complexity parameters between the transformed signal and the input signal, which may explain the improvement in the classification scores. SNESM could prove to be a useful time series signal processing system designed to improve accuracy in classification tasks.

## 2. Introduction

Nowadays, one can observe a drastic technological development and a real "explosion" of data created and accumulated through various activities. Scientists are beginning to pay attention to an effect that we are already slowly beginning to observe in the world, the so-called informational black hole. This concept is based on the fact that such a large amount of information is generated that we are physically unable to process it [1-4]. For this reason, vital information can get lost in a "black hole" overwhelmed by the commonly generated data noise. This should not come as a surprise at a time when we can have, among many others, a smart refrigerator, smart light bulbs, TV, telephone, washing machine, vacuum cleaner, curtains, heating, and assistant (e.g., Apple Siri, Google Assistant, Samsung Bixby) in our homes. These complex information collection and processing systems are known as the "Internet of Things" (IoT) [5-7]. Additionally, a huge portion of data is generated every second in cyberspace. We use social media, messengers, online stores, video players, etc. The data collected on each of these platforms is analyzed and profiled in order to be able to provide us with more personalized advertisements and to make a bigger profit for the service provider. Furthermore, it provides us with information published by our friends or groups to which we belong or subscribe, topics which attract our attention because of our aesthetic, intellectual, or political preferences [8]. The collection and analysis of those data translates into real-world effects in the form of monetary gains/losses [9,10], our choices as customers, and finally as voters (the case of Cambridge Analytica affecting the US presidential election and Brexit) [11].

To address the computational problems of the modern age, scientists are working to improve existing hardware and software technologies, as well as develop unconventional and/or hybrid approaches known as heteroic computing [12]. The growth of the first approach is commonly known to be subjected to Moore's law, whereas the second approach attempts to overcome its limitations through embedded multifunctionality and interaction with the environment [13]. Moore's law states that computational capabilities of conventional transistor technologies will double every few years. However, there is a lower limit to miniaturization due to (i) the granularity of matter and (ii) the effects of quantum tunneling of electrons through the gate of the transistor and (iii) heat management problems, which are the main reasons for looking for other computational technologies [14-19]. Second major issue is the so-called Von Neuman Bottleneck. It is a problem that limits the computing abilities of classic computers resulting from the separation of memory functions and processing that require communication between these components, which in turn limits the speed of the computing process. One of the novel technologies that aims to resolve some of modern computing problems is the application of memristors (and other memristive elements) and memristive circuits [20-24].

In a classic configuration, capacitive material (e.g. dielectric or semiconductor) is placed between metal electrode contacts in a sandwich-type system. Basically, memristive materials have two resistance states, called HRS (high-resistive state) and LRS (low-resistive state). By applying electrical pulses (or scans) of proper voltage, materials can be SET or RESET to LRS or HRS depending on the memristor type and its initial state. Due to some preservation of states after the turning off of the power source, resistive switching effect is hoped to be used as building block for novel non-volatile memory. In this configuration, it would be a new genre of information storage elements, called ReRAM (resistive, random access memory) [25], which would compete with currently available flash memory and some other continuously developed technologies (magnetoresistive random access memory, etc.). Nevertheless, the high demands on the energy efficiency of the switching, its statistical repeatability, and speed of operation make the new materials are still being sought to meet them. The application of novel elements, such as memristors, allows delegatement of some of the computational steps of artificial neural networks to be realized *in materia* [26].

Many scientists conduct intensive research that draws inspiration from biological neural structures in order to achieve more efficient computational structures, potentially of a universal nature. The high popularity is evident by an increasing number of publications appearing each year, scientific journals, and numerous conference meetings related to this subject. Interested readers are referred to recent reviews on the subject [27-30]. As can be seen, neuromorphic computing appears to be moving towards "conventional" computing, or at least towards special purpose computing modules [31]. Recurrent neural networks, which are especially good at representing dynamics of given input due to feedback loops present in the system, exhibit problems with costly learning process. To solve this problem, Jäger and Maas independently proposed the Echo State Network (ESN) [32] and the Liquid State Machine (LSM) approaches [33]. In their constructs, in contract with artificial neural networks, the information processing layer is not to be trained, only the readout layer is subjected to training procedures. Thus, they suggested the importance of the multidimensional, rich, and dynamic state space of the information processing layer [26,34]. Over time, both of these approaches to efficient training of recurrent neural networks were incorporated into a common conceptual framework named Reservoir Computing (RC) and the information processing layer was named the "reservoir" [35,36]. The reservoir in the RC paradigm describes a computational substrate capable of representing various inputs in a multidimensional configuration space of states, where computation is represented as a trajectory between successive states of the system in this space. Hence, as a proof of concept, RC has been implemented in simple systems as a bucket of water, where the data set with human speech was encoded as a series of water splashes [37]. Pictures of the perturbed water surface and used as basis for classification tasks. In our previous work we have shown that RC computing systems can be implemented even in such primitive setups as a doped cement and successfully used to classify simple signal according to its shape [38].

These unconventional computing systems must have several features in order to function properly, namely they must show appropriate internal dynamics, volatile memory, and the echo property of previous states. The RC circuits are based on: (i) a nonlinear element (e.g. a memristor) that also provides memory functions, (ii) an input layer that provides information/signal for processing, (iii) readout layer, and optionally (iv) a delayed feedback loop that provides internal system dynamics [39,40]. The readout layer is the only part of the system that needs to be „trained" and that is capable of sampling and representing the internal state of the reservoir. In other words, only simple model training is performed on the given features of signal transformed by the reservoir. In this case, no sophisticated ANN network is needed as readout, simple models such as linear regression, decision tree, etc. suffice. There is an obvious trade-off between the complexity of reservoir layer and the complexity of the readout process needed for a given classification/prediction task [41].

Single-node echo-state machine (SNESM) are novel RC systems that has only one physical node operating in the delayed feedback loop [26,42-45]. Basically, both the signal and the physical node state change each time the signal passes through the device in each cycle. The evolution of the loop signal can potentially improve clustering and classification capabilities of the RC system. Wlaźlak et. al. showed implementation of SNESM system based on a single hardware memristor for the signal classification task based on its amplitude [43,46]. In our previous work we have shown clustering of musical consonant/dissonant intervals through signal evolution in the software RC system based on the bridge synapse operating in the feedback loop [47]. Bridge synapse consists of four memristors and a differential amplifier. It was shown that bridge synapse is good at supporting lots of analog states, which can be beneficial from the perspective of rich state space needed for a good reservoir computing substrate.

Epilepsy is a disease that affects approximately 50 million people worldwide [48]. Epilepsy is diagnosed based on the appearance of seizures, which are defined as transient signs and/or symptoms resulting from abnormal activity of a population of brain nerve cells [49]. Therefore, epilepsy is considered a chronic brain disease characterized by an enduring disposition towards recurrent unprovoked seizures and by the neurobiological, cognitive, psychological, and social consequences of this condition [50]. Causes of epilepsy are different and include structural, genetic, infectious, metabolic, immune, and finally unknown factors [49]. There are several types of epilepsy that are classified on the basis of the type of seizure and pattern of brain activity [51] recorded with the use of non-invasive electroencephalography (EEG) through electrodes affixed to the scalp [52]. Currently seizures can be controlled and it is estimated that 70% of patients could live without seizures with appropriate medication [48]. However, the occurrence of seizures is unpredictable and, therefore, the development of methods of seizure detection is of great importance for a significant improvement in the quality of the patient, especially with a risk of sudden unexpected death in epilepsy (SUDEP) [53]. It should be noted that while in many cases seizures are associated with muscle activity, usually stiffening and then jerking of an arm or leg (tonic-clonic seizures), there are also types that are not associated with motor manifestations, for example, symptoms like hot and cold feelings, abnormal visual, olfactory, auditory, gustatory and somatic sensations, or seizures characterized by interruption of activities that occur suddenly with impairment of consciousness (absence seizures) [50]. Currently, EEG combined with video recordings are the most reliable methods for detecting epileptic seizures; however, they are only available in ambulatory monitoring. Thus, portable seizure detection devices have been proposed as an alternative solution [54].

As of today, deep learning methods allow classification of epilepsy using EEG signals with a precision of 99%, but those models require large datasets for training and classification tasks, and are cost ineffective [55]. Artificial data augmentation methods often must be used as a large amount of epilepsy seizure data is generally hard to obtain. One of the advantages of RC systems is their capability of operating with scarce, incomplete data employing only the simplest classificators (at the read-out layer) like linear regression, gradient descent, or decision trees. There is no need for multilayer artificial neural networks to train the SNESM read-out layer, as the reservoir performs data transformation that aims at simplifying classification task. The SNESM system gives additional benefit in the form of dataset extension. Each consecutive signal epoch is slightly different version of the base signal due to non-linear transformation and damping in a delay-based RC system. Echo of the primary signal may be easier to classify due to changes in its complexity and correlations between the parameters that describe it.

In this work, we present characteristics (noise characteristics and total harmonic distortion) of KNOWM memristors and bridge synapse based on them. Furthermore, we present analysis of reservoir layer and results of evolution of classification accuracy score of analyzed signal in the SNESM system based on hardware bridge synapse. For the classification task, we used a dataset containing cases of epilepsy, walking, running and sawing (overall 408 instances) collected from three different patients using a triaxial accelerometer attached to the dominating hands, which is a common technique of a non-invasive and non-disturbing way of monitoring motor activity [56-58]. On the contrary, other techniques, such as electroencephalography, electrocardiography, and electromyography, are impractical in daily life. The data for each patient were then analyzed separately. A scarce dataset is used on purpose, as with the RC approach, and the system could be calibrated individually for each patient in a cost-effective manner. Evolution of the signal in the system, especially damping, noise, and distortion accumulation, as well as nonlinear filtering influences improvement of a f1-score of classification of used signals. F1 score is a statistical analysis used to test the classification accuracy of the system [59]. For this data set we have found that echo of the base signal presented a better classification accuracy of epilepsy in relation to unprocessed waveforms. Furthermore, when epochs were analyzed collectively, further improvements of classification accuracy were obtained. In addition, changes in distribution and correlation between complexity parameters are shown for the input signal and transformed signal, which could explain changes in classification scores. This work poses as proof-of-concept that evolution of the signal in SNESM systems can highlight features of given signal or transform it in such a way that the overall classification of it can be improved in the case of scarce dataset and simple classification model.

## 2.1. KNOWM memristors

KNOWM Inc. memristors are semi-commercial memristive devices designed for scientific research [60]. It is very convenient at prototyping various memristive circuits. Several scientific works use KNOWM memristors in their research [61,62]. The researchers presented the possibility of constructing a hardware chaotic Chua oscillator without the need to use equivalent circuits [63]. The other group showed the characteristics of the impulse response and the possibility of coupling the operation of KNOWM memristors with an ultrasonic transducer [64]. These memristors are good candidates for the construction of hardware neural networks based on thin-film nanomaterials because of their good and mostly repeatable performance and simplicity in the prototyping of networks with various structures. The schematic layer diagram of the device is shown in Fig. 1.

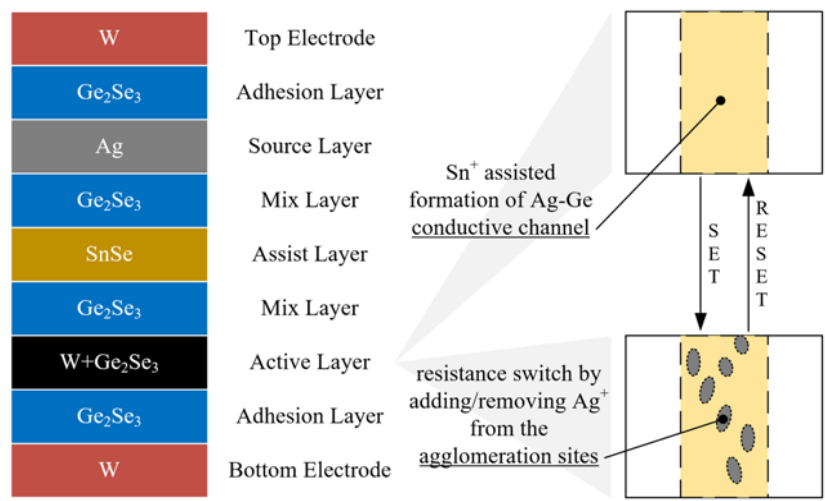

*Figure 1. Layer diagram of the thin-film stack of nanomaterials in KNOWM memristive devices and depiction of the switching mechanism. Reprinted from Ref. [65].*

The mechanism of operation of KNOWM memristors proposed by the manufacturers is as follows. After manufacture, the devices are in the HRS state (MΩ – GΩ range). When the device is put into operation for the first time, a so-called self-directed channel is formed when a positive potential to the upper electrode. In this case, the required potential is the same as that used during normal operation of the device. Applying a positive potential to the memristor generates Sn ions from the SnSe layer and forces them into the active $Ge_2Se_3$ layer (here W-doped). The Sn ions are intended to facilitate the transport of Ag to the active layer at the Ge-Ge binding sites. This occurs due to the energetically beneficial process in which the electrons entering the active layer ($Ge_2Se_3$) from the negatively polarized bottom electrode (with the simultaneous formation of Sn ions from the SnSe layer) are strongly localized around Ge-Ge dimers. During this reaction, the glassy network of $Ge_2Se_3$ is distorted, creating an "opening" near the Ge-Ge sites. That opening allows the $Ag^+$ ions to access possible Ag-Ge binding site more easily and become preferred conductive channels in the active layer for $Ag^+$ movement during device operation. As a result, the Sn ions facilitate the energetically favorable Ag/Ge substitution reaction at the Ge-Ge bond. This self-directed conductive channel is a result of the natural structure of the glass and follows the position of the initial Ge-Ge dimers in the glass. Since Ag tends to agglomerate with other Ag atoms, the Ag-Ge substitution sites may favor Ag agglomeration in the glass. Accordingly, the resistance of the device is altered by adding or removing Ag from agglomeration sites within this *in situ* generated path. It is anticipated that conduction may occur between clusters of Ag agglomerations in the glass. Therefore, the path need not be composed of conductive metallic filaments that include two opposing electrodes, as in CBRAM devices. In the case of KNOWM memristors, the device resistance is dictated by the Ag concentration in a given place of the agglomeration and the distance between successive places of the Ag agglomeration. Resistance regulation is carried out by the movement of $Ag^+$ ions between the device layers, forced by the application of a positive or negative electric potential [60].

## 3. Results
### 3.1. Large-amplitude sinusoidal voltammetry

LASV (large amplitude sinusoidal voltage) is a classic electrochemical measurement technique, used here used to characterize memristive devices, to determine threshold of their switching potential, speed of operation, and its repeatability. The probing signal is a simple sinusoidal wave of given frequency and amplitude. The scans were performed in the potential window for which switching from ON/OFF (or Low Resistive State/High Resistive State) states is repeatable, in order to determine the optimal "SET" and "RESET".

To prevent memristors from being burnt, measurements were made with a series 5 kΩ resistor in the circuit to limit the current. Too high currents (above 1mA) irreversibly damage KNOWM memristors, irreversibly switching them to LRS. On the other hand, too low currents prevent the system from switching to the LRS state. This was observed during memristor tests in which a 20 kΩ resistor was used. For a single chip (16 memristors), for a 20 kΩ resistor, 4 out of 16 memristors did not switch to the LRS state. In the case of the 5 kΩ resistor, all memristors were switching and had good cycling repeatability.

In this study, LASV measurements were performed at 19 different scan speeds ranging from 1 to 100 Hz (example in Fig. 2a). The results are shown in Figs. S1 and S2. In this frequency range, KNOWM memristors operate reliably, as producers show in the data sheet, that they can operate even at 1000 Hz. LASV measurements were performed for three scenarios: for a single memristor and for a bridge synapse without and with a differential amplifier.

### 3.2. Total Harmonic Distortion

On the basis of the results from LASV measurements, an analysis of the frequency distortion was performed. This was done by determining the total harmonic distortion (*THD*) parameter. It tells us about the contribution of higher harmonics of the total signal to the fundamental frequency of the signal and can thus be considered as a simple experimental index of device linearity [66]. It is especially useful in the characterization of memristive devices, because it gives a clear distinction between unipolar and bipolar memristors [67]. To determine THD, Fourier spectra of the analyzed signals are needed (an example spectrum is shown in Fig. 2b). Additionally, Fig. S3 shows a Fourier transform of an exemplary driving signal. Based on the calculated Fourier transform for all frequencies analyzed, the fundamental *THD$_F$* was determined on the basis of the following formula (1):

$$THD_F = \frac{\sqrt{V_2^2 + V_3^2 + V_4^2 + \cdots}}{V_1}, \tag{1}$$

where *V$_n$* is the *nth* amplitude of the frequency of successive higher-harmonic peaks observed in the Fourier spectra. The THD parameter in its fundamental form can take values greater than 100%, hence the following correction can be applied as normalization (2):

$$THD_R = \frac{THD_F}{\sqrt{1 + THD_F^2}}, \tag{2}$$

where *R* in *THD$_R$* stands for "root mean square".

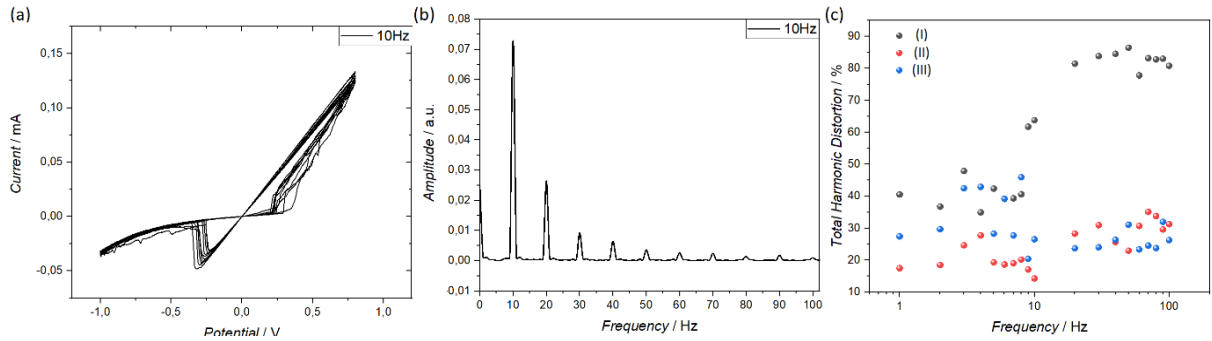

*Figure 2. Exemplary LASV signal measured at 10 Hz (a) and Fourier spectra (with rectangular apodizing function window) based on it (b). Total harmonic distortion determined on the basis of Fourier transform, for frequencies ranging from 1 to 100 Hz (c). Results presented for single memristor (I), bridge synapse without differential amplifier (II), and with differential amplifier (III).*

Fourier spectra clearly indicate bipolar character of the studied memristors, as all even and odd harmonics are equally well represented in the spectra (Fig. 2b) [67]. The results in Figure 2c show that higher $THD_R$ occurs at frequencies above 10 Hz for a single KNOWM memristor, while for bridge synapse circuits, $THD_R$ is generally constant (apart from several cases in lower frequency range shown in Figure 2c). This is a marked difference between KNOWM memristors and reported theoretical models, where the introduction of a differential amplifier into the Wien-like bridge resulted in a much increased $THD$ [47].

### 3.3. Noise Analysis

Chronoamperometric methods were used to characterize the noise present in KNOWM devices in the ±1 V range. A set of potentials was applied step manner (step size 100 mV) to the device, and the response was measured for 100 s. The average values of the measured current at each step during the measurement are shown in Fig. 3a. These values resemble the classic hysteresis loop observed during LASV measurements. Presented in the time domain, the alternating switching of the four memristors in the bridge synapse (without differential amplifier) can be observed in Fig. 3b. If plotted in I-V (or V-V) domain, the result from Fig. 3b would resemble the one on the Fig. 3a. In the next step, a Fourier transform was performed to each voltage step to calculate the noise spectra (Fig. 3c).

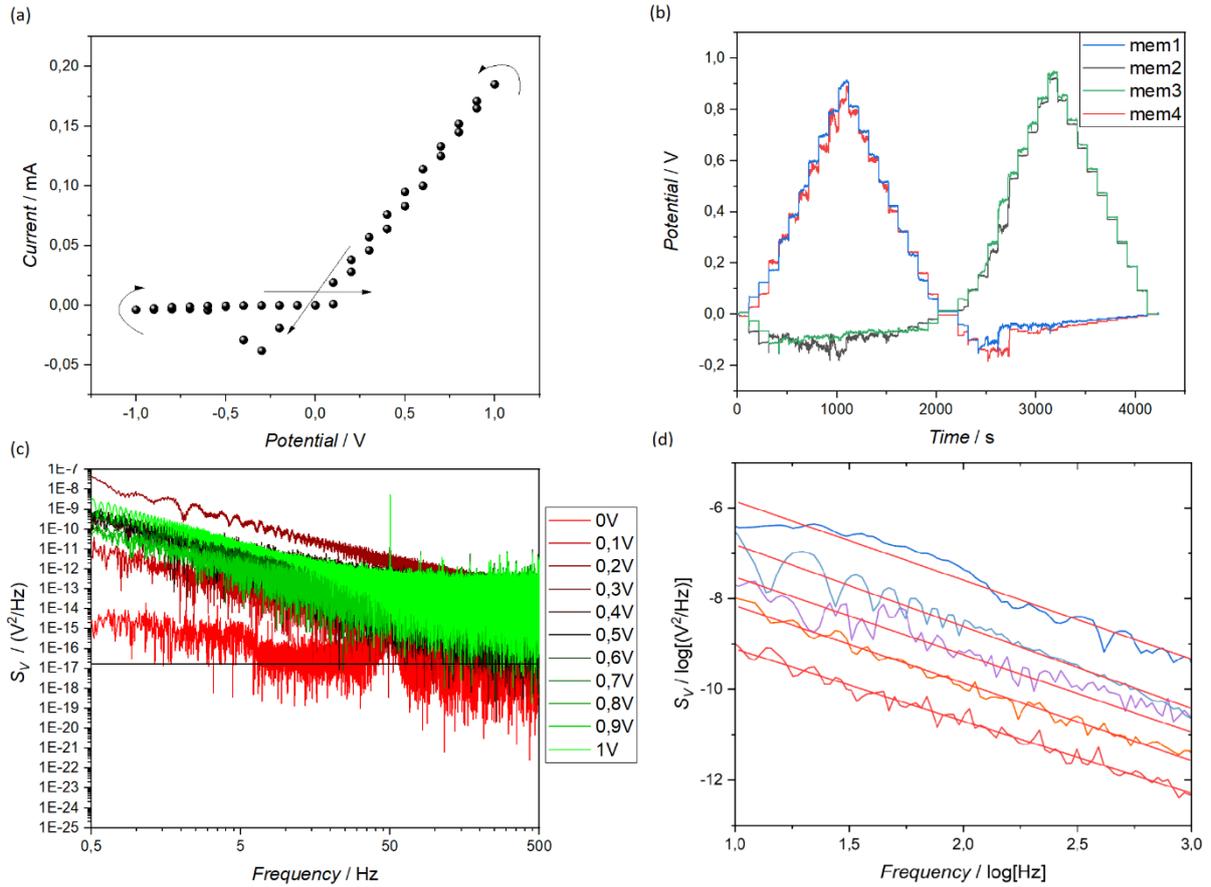

*Figure 3. Average current values for the noise measured in the potential window ± 1V for 100 s for each step of 100mV. The arrows indicate the direction of the measurement. Measurement performed for a single memristor (a). Voltage difference recorded between individual elements of the bridge synapse during noise recording measurement (b). Noise measured for KNOWM memristors presented in frequency domain, for potentials from 0V to 1V. The black line indicates the thermal base floor measured at 0V (c). An example of fitting a line to the several filtered noise spectra (d).*

The slope of the noise spectra was determined in order to compare the noise character at different potentials (Fig. 3d). Due to the fact that the results are presented on a logarithmic scale, in order to reliably fit the lines, it is necessary to interpolate recorded traces, and reduce point density to achieve uniform data point distribution in all frequency ranges.

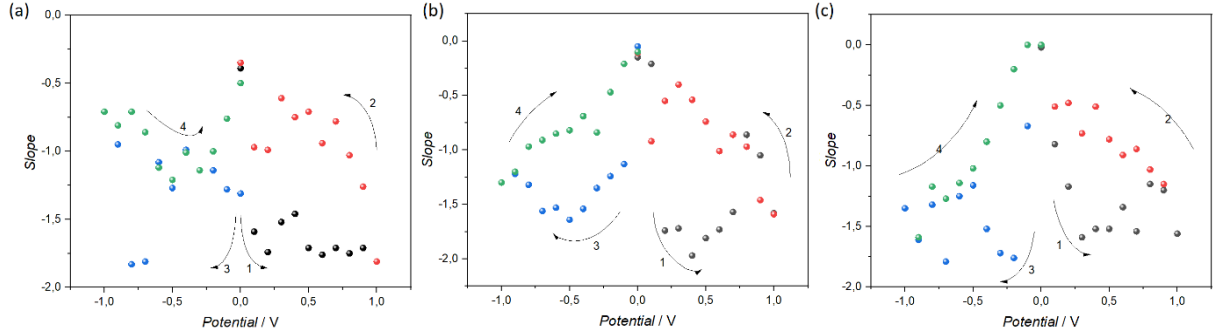

*Figure 4. Collected slope values of noise spectra at different DC potentials. Arrows indicate direction of the experiment. Results for single KNOWM memristor (a), synaptic bridge (b), and a circuit comprising a bridge synapse and a differential amplifier (c).*

Generally, the noise spectra observed in the case of KNOWM devices are of red/brown character [68], described by the power law of the form (3):

$$S_V(f) \propto \frac{1}{f^\gamma} \qquad (3)$$

where $S_V S(f)$ is the spectral power density of the signal and γ is a parameter that reflects the noise type (0 for white noise, -1 for pink one, and -2 for the brown noise. In the studies cases, we can see that the thermal noise of the KNOWM memristor evolved from white through pink to brown noise, depending on the applied potential and its history, as the noise exponent for all studied circuits follows the hysteresis loop, however the point s are a bit more scattered. During the stepwise voltage scans, the γ parameter changed from close to zero at very low voltages down to -2 for voltages higher than ca. 0.5 V. At ca. 1 V (the point of transition from HRS to LRS), the noise parameter gradually increased to zero with decreasing voltage (Fig. 4a). This behavior indicates that the charge carrier dynamics in multilayer KNOWM memristors is complex and voltage-dependent. Voltage dependence may suggest electron-hopping-type conductivity [69].

Santa et al. suggest internal fluctuations between metastable atomic positions as one of the possible mechanisms of observing 1/f type noise in Ag filament-based memristive materials. Furthermore, redox exchange is another possible mechanism, which in our studies would originate from the binding of $Ag^+$ ions to the Ge-Ge sites during the switching. Diffusion of Ag + ions inside Ag clusters may also have an influence on registered noise [70,71]. This suggestion is consistent with the observed characteristics. In the HRS conductivity is dominated by the hopping impurity mechanism, and results in low γ values ($\gamma \leq -1$), which depend on applied potential only weakly. In LRS state, the γ values are significantly higher (γ ≥ −1), and increase with decreasing voltage across the device. It is associated with mixed conductivity type/; Ohmic via filaments (or silver agglomerates) and impurity hopping in the bulk of the material.

The observable noise signature demonstrates the presence of multiple conductivity channels that form a complex network within a single device. The source of the noise can thus be described as an internal switching dynamics of an atomic switch network with a small world architecture [36].

It is in accordance with the literature studying 1/f noise present in memristive devices. For a bridge synapse (Figs. 4b and 4c), white noise is observed at potential close to 0V. The most symmetric response is observed for the bridge synapse without a differential amplifier (Fig. 4b), as this configuration is symmetrical in design. Symmetric switching of the KNOWM memristors in the bridge synapse is shown in Fig. 3b.

### 3.4. Analysis of simple signal transformation in SNESM

As part of the initial characterization of signal transformation in the SNESM system, simple signals of various shapes and frequencies were applied to the reservoir. Frequency ranged from 1 to 10Hz in 1Hz steps and from 10Hz to 100Hz in 10Hz steps were used. The delay time of the returned signal was determined as 20% of the signal packet duration, which provided an equal relative distance of repeated packets between different frequencies (an example shown in Fig. S4).

To analyze differences in signal damping, we fitted an exponential decay function to the maximal potential values of all signal packages. The time constant of the fitted exponent was collected for the frequencies and signal shapes (results shown in Fig. 5).

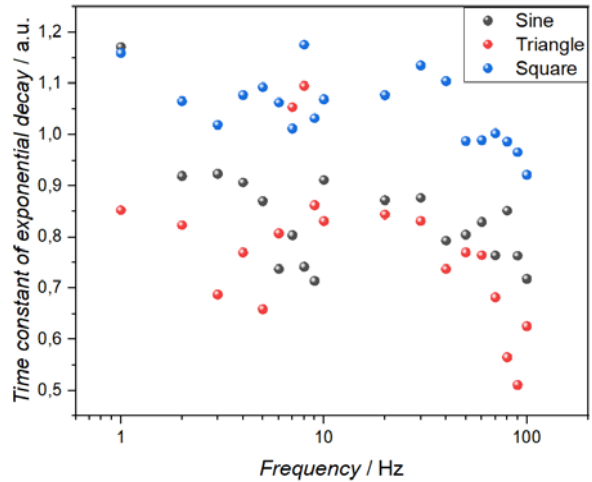

*Figure 5. Collection of the values of exponential decay time constant for three signal shapes. Signal measured in SNESM with bridge synapse as computational substrate.*

Results show that for nearly all studied frequencies, clear separation of time constant can be noticed, however a substantial overlap is also present in some frequency ranges (ca. 5-10 and 50-100 Hz). That result further suggests that signals of various shapes and frequencies can be represented in the SNESM system differently, possibly enabling their classification. Furthermore, signals of different shapes are filtered in a different way by a reservoir. Generally, triangular-shaped waves were damped the most, whereas square waves were passing with less attenuation. This result supports the presence of two fundamental reservoir requirements: separation and generalization properties – signals of different shapes are separated by the system, and this separation is generalized across different frequencies (with slight deviations at 1, 5 and 6Hz). Finally, the memory of KNOWM memristor (in addition to its resistive switching properties) can be represented as changes in the noise characteristics, which is sensitive to the system history, which is another important feature sought in the RC paradigm.

### 3.5. Epilepsy detection

The classification possibilities and the influence of the RC system on signal evolution were tested in the recognition of signals representing an epilepsy attack in patients. The dataset contains results for three patients, with 136 instances per patient. The classes included in the dataset are "walking", "running", „sawing", and "epileptic seizure". The data set contains time series of equal length.

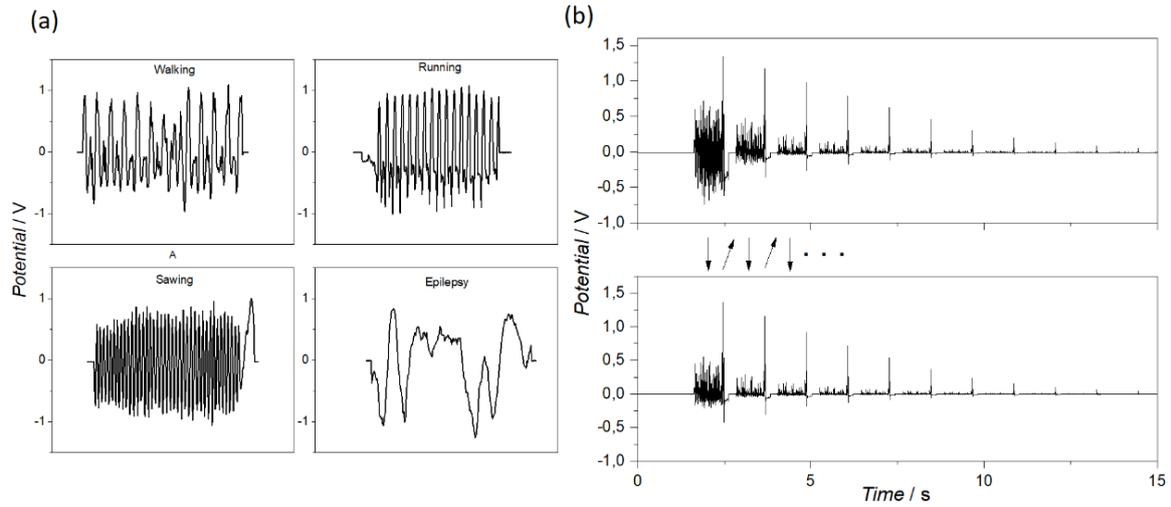

*Figure 6. Examples of four classes of signals from the dataset (after standardization) (a). Evolution of the signal in the SNESM system (b). First, the driving signal passes through the system, then is delayed and feed back again (arrows indicate direction of data flow).*

Due to the nonlinear nature of the operation of memristors, which translates into non-linearity in the transformation of signals through the entire bridge synapse, strong filtering of negative part of signals can be observed (Fig. 6b). Individual signal epochs exhibit stationary character which was confirmed with the augmented Dickey-Fuller as well as the Kwiatkowski-Phillips-Schmidt-Shin test. Subsequently, 11 parameters characterizing the complexity of the signal were calculated, which served as features for the ML model training. An example of the correlation analysis (Pearson's correlation parameter) of the calculated features can be seen on Fig. 7. The presence of positive and negative correlations in the dataset is desirable, potentially contributing to a good classification accuracy. The calculated parameters were then used in training/testing a decision tree classification algorithm.

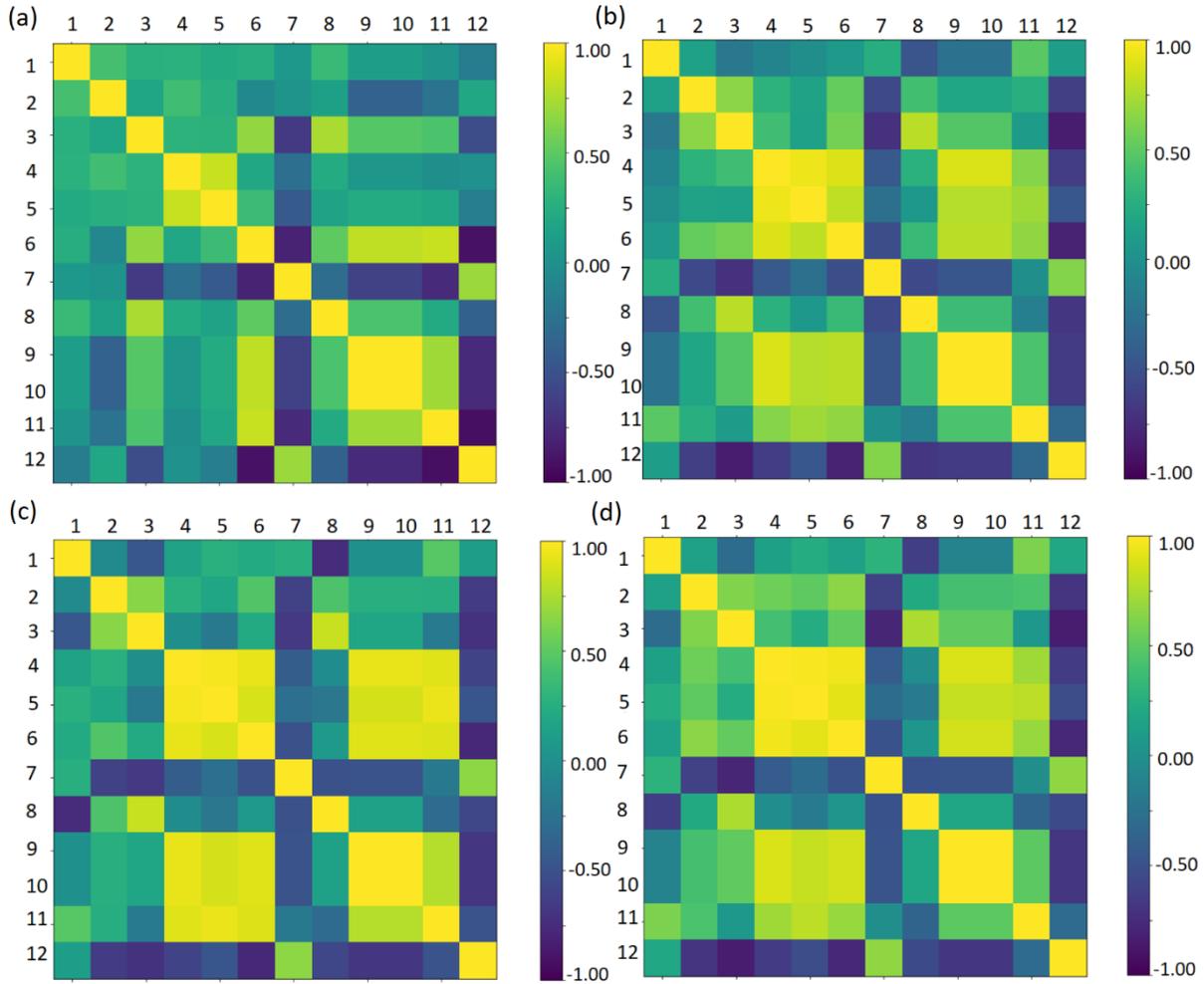

*Figure 7. Exemplary result of Pearson's correlation for the "features" of the signal used by the reservoir system (a). The numbers correspond to the following parameters: 1- 'Permutation entropy', 2- 'Spectral entropy', 3- 'Singular value decomposition entropy', 4-'Approximate entropy', 5-'Sample entropy', 6-'Hjorth mobility ', 7 -'Hjorth complexity ', 8-'Number of zero crossings', 9-'Petrosian Fractal Dimension ', 10-Katz Fractal Dimension', 11-'Higuchi Fractal dimension ', 12-'Detrendend Fluctuational Analysis. Results for unprocessed input data (patient no. 1) (a) data transformed with SNESM: all 5 epochs (b), only second epoch (c), only $5^{th}$ epoch (d).*

The analysis was performed for a maximum of 7 epochs of the signal (an increasing part of the signal was filtered in further epochs). Analysis was performed for each individual epoch (Fig. 8a) as well as for several epochs at the same time (Fig. 8b). The classic statistical parameter was used to determine the accuracy of the classification performed by the SNESM. It is the harmonic mean of the 'precision' (known as a positive predictive value) and "recall" (also known as sensitivity) classification parameters. Precision is the fraction of relevant instances (target class, "Epilepsy" in this study) among the other instances, whereas recall is the fraction of the relevant instances that have been classified correctly among other instances from the same class that have been omitted. F1 scores are biased to the lowest value of each precision and recall, so when the F1 score increases, both precision and recall parameters will be increased and balanced. In Fig. 8 are presented results of the classification accuracy for increasing number of consecutive signal epochs.

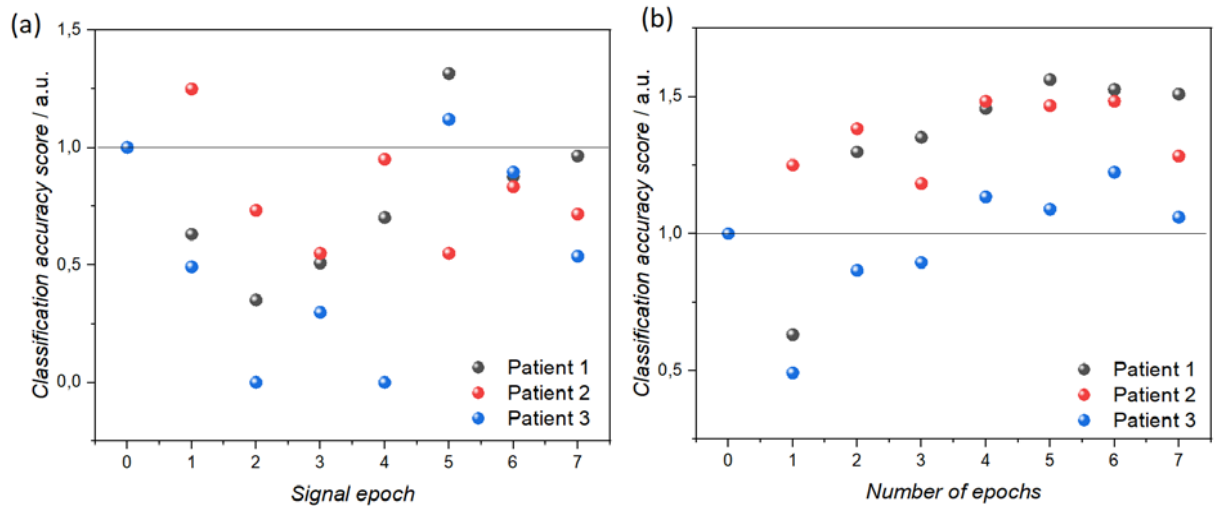

*Figure 8. Result of classification precision scores for consecutive signal epochs (a) and for several epochs at once (b). The number '0' symbolizes unprocessed input data that serve as a reference.*

The results obtained show that after several cycles, system exhibits an improvement of classification accuracy in comparison to unprocessed data. When the data were analyzed collectively, several epochs at once, further improvements in accuracy were obtained. To illustrate and compare the operation of the SNESM system for different patients, results were normalized in relation to the accuracy of unprocessed data, representing '1'. This method of data presentation allows for a simpler comparison of data sets from different patients for which the initial classification accuracy serving as a reference may differ. Apart from the case of patient no. 2, the first signal epoch actually worsens the classification accuracy. However, taking into account first and second signal epoch, accuracy is improved for all patients even though for individual epochs the accuracy was worse than reference. This effect may arise due to SNESM capabilities of data expansion and data transformation.

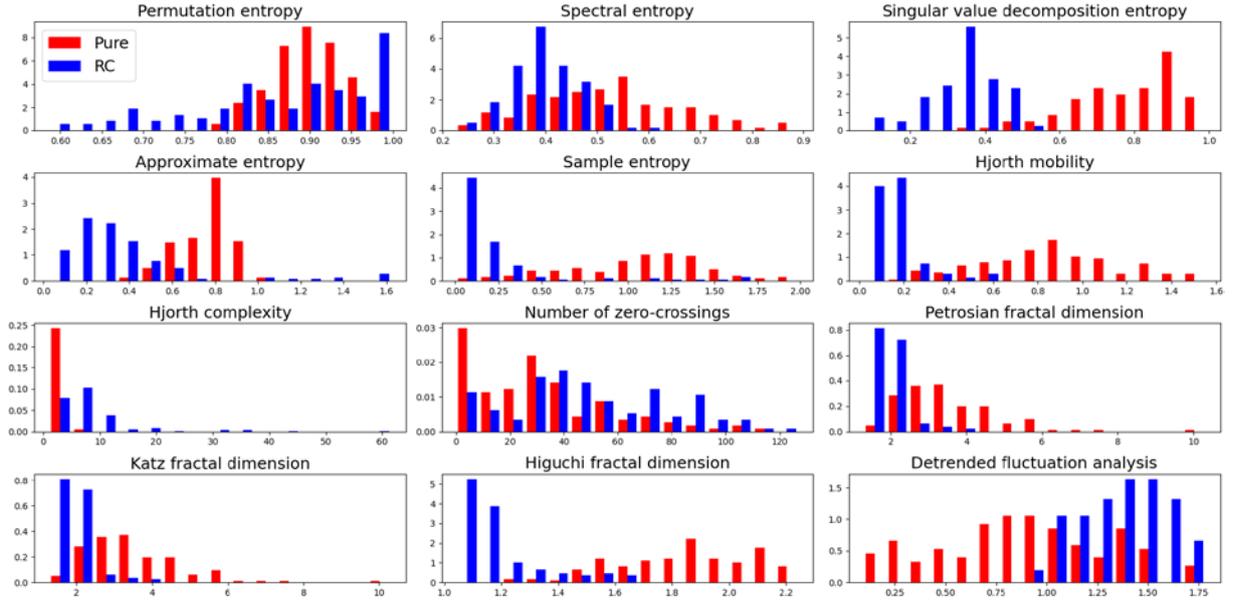

*Figure 9. Comparison of histograms for the complexity parameters of the unprocessed signal ("Pure", colored red) and the transformed signal in the reservoir computing system ("RC", colored blue) for the patient no. 1 (Epoch no. 5 for RC).*

To inspect changes in complexity of the transformed signal histograms of two cases were compared – unprocessed signal and epoch no. 5 of this signal transformed by the SNESM for patient no. 1. Those signals were chosen as in this case, the highest improvement of classification accuracy was observed for a single signal epoch. Almost all of the calculated complexity parameters show distinct changes in their distribution, which can be one of the main reasons for improvement in the classification task in question. Not only those distributions changed, but also correlations between individual parameters have changed, where for transformed signal, more positive correlations are present for parameters which previously exhibited weak correlation (Figs. 7a, b). At the same time, for the 2$^{nd}$ epoch case, the highest number of extreme correlation values can be observed, where for this case the classification accuracy result was the worst. The correlation results for the fifth epoch (Fig. 7d) are more similar to the correlation results for the cumulative five epochs (Fig. 7b), which may suggest that there are some optimal parameter correlation values that maximize the final ability of the system to accurately classify data.

## 4. Summary

In this work we have presented measurements and analysis of $THD_R$ and noise in KNOWM memristors and bridge synapse based on them. Furthermore, the bridge synapse was incorporated into SNESM as a computational node. Analysis was performed on damping of signal of three standard different shapes (sin, square, and triangle) based on time constants of fitted exponential decay curves. After establishing rich internal dynamics, memory, separability, and generalization properties, the SNESM system was tested as a tool to improve the classification of epilepsy. The system was tested under the constraints of scarce datasets (data from three patients classified in parallel) and using a simple classification model (decide tree). The results show an improvement in classification accuracy for several transformed signal epochs. Further improvements were achieved when signal epochs were analyzed collectively. This result shows that by transforming and filtering data in the SNESM system, certain properties of a given signal are amplified and others weakened, which is represented by different values of its non-linear dynamics parameters and changes of their correlation.

Nonlinear transformation, filtering of the signal, and its evolution in the feedback loop have a positive effect on the overall classification accuracy score. Results of similar nature were shown in one of our previous works, where signal filtering through an artificial neuron helped to improve overall data separability [72]. As was shown before, bridge synapse exhibits rich dynamics represented in several ways: in its noise characteristics, $THD_R$ and time constant of exponential decay. As various RC systems aim to be more general in their computation capabilities, those facts may serve as an indication that this SNESM based RC can be a good computational system for other classification tasks. Results of characterization of KNOWM memristors and bridge synapse-based SNESM system suggests the possibility of employing such systems for other computational and/or classification/prediction tasks, with scarce datasets and simple classification models.

## 5.  Experimental

Three different circuits used in this study are shown in Fig. 10. BioLogic Sp-150 was used to measure LASV and chronoamperometric noise measurements. ADA-24 PicoLog was used to probe individual KNOWM memristors in the bridge synapse. An analog channel of BioLogic was used to record signals of simple shapes (sin, triangle, and square). Fitting of exponential decay curves and collection of their time constants were performed using Python (Pandas and SciPy packages). OriginPro was used in calculations of the fast Fourier Transform for THD and noise analysis.

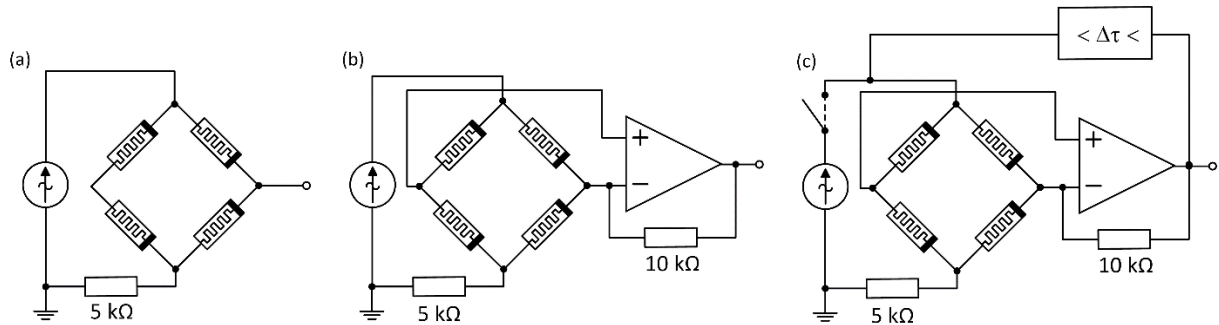

Figure 10. Schematics of the systems studied. Bridge synapse without (a), with differential amplifier (b), and incorporated into SNESM system (c).

The dataset for the classification task was downloaded from the "UEA & UCR Time Series Classification Repository" (https://www.timeseriesclassification.com/description.php?Dataset=Epilepsy). Firstly, the data set was standardized using Python. The arbitrary function generator TTI TG5012A was used to apply the signals to the RC system. Data were uploaded from a flash drive in bits of four (maximum TTI TG5012A memory). Each waveform required additional scaling to be in the $1V_{pp}$ range due to the observed lack of consistency in the amplitudes of the signals generated by the TTI TG5012A. Before each measurement, an Agilent oscilloscope was used to fine-scale the signal. The Rigol DG4062 generator was used as the arbitrary trigger for the TTI TG5012A signal source. Furthermore, the TTI WA301 amplifier was used as a current buffer. The bridge synapse includes the TL082 operational amplifier (powered by a 9 V battery) in differential mode - the negative input of the operational amplifier has an additional feedback from the amplifier output connected through a 10 kΩ resistor. The analog channel of the Biologic SP-150 potentiostat was used to record the data. Custom-made delay line (Chip, Poland) was used in the setup. The generator signal was first sent to delay, then to the Kacper-01 analog potentiostat (Instytut Fotonowy, Poland), which transmitted the signal to the synaptic bridge. The differential amplifier passed the signal back to the delay (via the TTI amplifier). The delay signal was recorded, delayed, attenuated (~ 10%) and finally circulated back to the system until fully suppressed.

The recorded signal was then analyzed using Python. To analyze reservoir properties, an exponential decay function was fitted to maximal amplitudes of given repeated signal instances. Time constants of exponential decay were collected and presented in Fig. 5. The extraction of signal complexity features was performed using the AntroPy Python package. A short description of the calculated complexity parameters can be found in the SI. The parameters with appropriate labels were passed through a simple machine learning pipeline (using Python Scikit learn package). Initially, general descriptions of the data (histograms and correlation matrix) were generated. A 10-fold cross-validation was used to divide the data set into training and testing sets. It involves dividing the data set into 10 parts, using 9 parts as a training set and 1 part as a testing set. This procedure is then repeated 10 times so that each piece of data is used as both a training and a testing set. As a classification model, the Decision Tree Classifier implemented in scikit-learn Python package was used. The complexity parameters of the unprocessed signal were used as a reference. Model training was performed on an unprocessed signal and for each consecutive delay signal epoch. In the next step, an increasing number of signal epochs were collectively used in training/testing the model.

## 6. Acknowledgements

The authors acknowledge the financial support from the Polish National Science Center within the OPUS (grant agreement No. UMO-2020/37/B/ST5/00663) and PRELUDIUM (grant agreement No. UMO-2018/31/N/ST5/03112) projects. DP has been partly supported by the EU project POWR.03.02.00-00-I004/16.